\documentclass[pre,aps,amsfonts,amssymb,amsmath,graphicx,showpacs,twocolumn,epsfig,amscd,ulem]{revtex4}

\usepackage{graphicx}

\def\bi{\begin{itemize}}

\def\ei{\end{itemize}}
\def\be{\begin{equation}}
\def\ee{\end{equation}}
\def\bea{\begin{eqnarray}}
\def\eea{\end{eqnarray}}

\DeclareTextSymbol{\degre}{OT1}{23}

\begin{document}

\author{N. Huang,$^{1}$ G. Ovarlez,$^{2}$  F. Bertrand,$^{2}$ S. Rodts,$^{2}$
P. Coussot,$^{2}$ Daniel Bonn$^{1, 3}
$ }

\affiliation{$^{1}$Laboratoire de Physique Statistique, UMR 8550 CNRS
associated
with University Paris 6 and Paris 7,\\ \'Ecole Normale Sup\'erieure,
24, rue Lhomond, 75231 Paris Cedex 05, France }

\affiliation{$^{2}$Laboratoire des Mat\'eriaux et Structures du G\'enie
Civil, UMR 113 LCPC-ENPC-CNRS,\\ 2, all\'ee
Kepler, 77420 Champs-sur-Marne, France }

\affiliation{$^{3}$Van der Waals-Zeeman institute, University of Amsterdam,\\
Valckenierstraat 65, 1018 XE Amsterdam, the Netherlands }

\title{Flow of wet granular materials}
\date{\today}
\begin{abstract}
The transition from frictional to lubricated flow of a
dense suspension of non-Brownian particles is studied. The
pertinent parameter characterizing this transition is the Leighton
number $Le = \frac{\eta_s \dot{\gamma}}{\sigma}$, which represents
the ratio of lubrication to frictional forces. The Leighton number
$Le$ defines a critical shear rate below which no
steady flow without localization exists. In the
frictional regime the shear flow is localized. The
lubricated regime is not simply viscous: the ratio of shear to
normal stresses remains constant, as in the frictional regime;
moreover the velocity profile has a single
universal form in both frictional and lubricated regimes. Finally,
a discrepancy between local and global measurements
of viscosity is identified, which suggests inhomogeneity of the material under flow.
\end{abstract}
\pacs{83.80.Hj,83.60.-a,47.55.Kf} \maketitle

In recent years, the flow behavior of granular matter has been the subject of considerable
controversy.
Simple questions such as 
whether viscosity can be properly defined
for granular systems are still the subject of hot debate
\cite{Gdrmidi2004, Jaeger1996}. A notion of viscosity is
necessary for many applications to predict the
resistance to flow. Perhaps even more crucial, in view of its
importance in geophysics and civil engineering,
is the resistance to flow of wet granular materials. Nevertheless, the
number of studies on wet granular matter is almost negligible compared to that
for sand.
The usual picture is that for an interstitial fluid at low viscosity,
the material behaves similarly to dry granular systems
(with frictional contacts between the grains),
and for an interstitial fluid at high viscosity
the material behaves similarly to viscously dominated systems
(with lubricated contacts between the grains)
\cite{Bagnold1954, Prasad1995, Ancey1999, Coussot1999}.

If this picture were true, it is of
importance to investigate what parameters determine the
frictional to viscous transition, and whether a viscosity can be
defined.

The purpose of this Letter is to answer some of these questions. We find
that the transition between frictional and lubricated flow regimes
is characterized by a
viscosity bifurcation \cite{Coussot2002, Dacruz2002}, and is governed by the Leighton number
$Le = \frac{\eta_s \dot{\gamma}}{\sigma}$, with $\eta_s$ the
interstitial fluid viscosity, $\dot{\gamma}$ the shear rate and
$\sigma$ the total shear stress. For a given interstitial fluid,
$Le$ fixes the critical shear rate
$\dot{\gamma_c}$ at which the transition
between the two regimes occurs.
In the frictional regime, 
no steady flow exists without localization.
We also find that the viscous regime is
similar to the granular flow regime posing the same problem for the definition of the viscosity as for dry granular matter \cite{Gdrmidi2004, Jaeger1996}.


We study the rheological behavior of a paste (a dense suspension)
composed of non-Brownian monodisperse spherical particles immersed
in a Newtonian fluid of the same density at a volume fraction
$\phi$ of 60\%. The particles are
monodisperse spherical polystyrene beads (diameter 0.29 $\pm$
0.03 mm, density 1.04 g.cm$^{-3}$). To avoid
sedimentation or creaming effects we match the density of the
interstitial fluid with the density of the particles. Adequately
salted water ($\eta_s = 1$ mPa.s) and Rhodorsil 10646 silicone oil
($\eta_s = 2300$ mPa.s) are perfectly density matched. The oil is
miscible with water: we mix it with salted water in order to have
an isodense liquid, whose viscosity varies between 1 and 2300
mPa.s. These experiments are completed with pastes prepared with
silicone oils of variable viscosities \cite{silicone}.
The rheology is done
with a vane-in-cup geometry on a commercial rheometer
that imposes either stress or shear rate. The
vane geometry we use is equivalent to a cylinder (diameter 16 mm,
height 52 mm) with a rough lateral surface on the scale of the
beads, reducing the slipping of granular materials
\cite{Ancey1999}. For the same
reason, the inside of the cup (diameter 26 mm) is covered with the
granular particles using double-sided  adhesive tape.

\begin{figure}
\includegraphics[width=9cm]{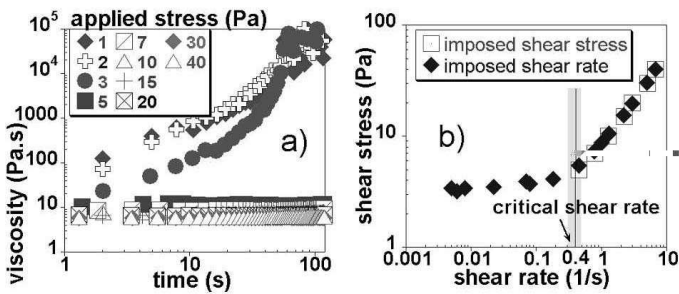}
\caption{(a) Viscosity bifurcation : under an imposed stress the
viscosity either grows in time, or decreases; therefore the steady
state viscosity jumps to infinity at a critical
stress. This allows us to define both the critical stress and the
critical shear rate. Before each experiment, the material is
presheared during 30 s at 30 s$^{-1}$ to obtain a reproducible
initial state. (b) Flow curve (shear stress vs. shear rate) for 20
mPa.s silicone oil, both at an imposed macroscopic shear stress
and shear rate. The shadow zone is the statistical error bar.}\label{fig1}
\end{figure}

In the first set of experiments, we use a 20 mPa.s oil as the
interstitial fluid. Fig. \ref{fig1}a shows the evolution of
viscosity with time, for different applied stresses. We find that
for stresses smaller than a critical stress $\sigma_c$ = 5 $\pm$ 3
Pa, the viscosity of the sample increases in time until the flow
is halted altogether,
which corresponds to an inifinite steady-state viscosity \cite{infinite_viscosity}.
For a stress only slightly above $\sigma_c$, the
viscosity decreases with time; for long times it reaches a (low)
steady state value $\eta_0$. This implies that at the critical
stress, the steady-state viscosity jumps
from an infinite to a finite and low value at $\sigma_c$. This behavior implies
the existence of a bifurcation of the viscosity,
which abruptly passes from a flowing state above $\sigma_c$ to a
jammed state below $\sigma_c$, as
observed previously for yield stress fluids and dry granular
materials \cite{Coussot2002, Dacruz2002}. Apart from a
critical stress, the bifurcation also identifies a
critical shear rate $\dot{\gamma}_c$ = 0.4 $\pm$ 0.1 s$^{-1}$, beyond
which steady flows are possible under an imposed
stress. If shear rate rather
than the stress is imposed, below $\dot{\gamma}_c$ the measured
shear stress is almost independent of the shear rate, which is the
hallmark of quasistatic granular (frictional) flow
\cite{Gdrmidi2004, Ancey1999}. For high shear rates
($\dot{\gamma}>\dot{\gamma_c}$), stress increases linearly
with increasing shear rate, as for a viscous fluid (Fig.
 \ref{fig1}b). These observations allow us to
identify the transition between two different
flow regimes.

\begin{figure}
\includegraphics[width=5cm]{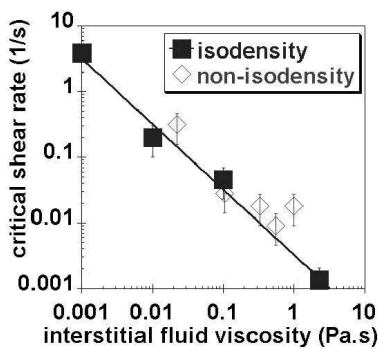}
\caption{Critical shear rate vs. interstitial fluid viscosity, for
polystyrene beads in a Newtonian liquid. The line is a power-law
fit to the isodensity values with a slope of -0.99 $\pm$
0.10.}\label{fig2}
\end{figure}

To find out what determines the transition between the two
regimes, we vary the viscosity of the interstitial fluid. As
before, we determine the critical stress and critical shear rate
are determined from the
viscosity bifurcation. Fig. \ref{fig2} summarizes the main
results. The critical shear rate $\dot{\gamma_c}$ is inversely  proportional to the
viscosity of the interstitial fluid, whereas
the critical shear stress $\sigma_c$ turns out to be
constant (within experimental uncertainty).
The former result is
consistent with that of Prasad and Kyt\"{o}maa \cite{Prasad1995}
and Ancey and Coussot \cite{Ancey1999}, who only used
two fluids of different viscosities.

We conclude from these data that the pertinent parameter
characterizing the transition is the Leighton number $Le =
\frac{\eta_s \dot{\gamma}}{\sigma}$, which represents the ratio of
lubrication to frictional forces. Ancey and Coussot previously
reported the inverse proportionality with respect to the frictional forces
\cite{Ancey1999, Coussot1999}. By varying the
lubrication forces, we conclude that the
transition is entirely characterized by the Leighton number.
The critical stress $\sigma_c$ does not vary significantly
($\sigma_c$ = 5 $\pm$ 3 Pa), so that $Le_c$ has the same order of magnitude
for $\eta_s$ varying from 10$^{-3}$ to 2.3 Pa.s
($Le_c \approx (7 \pm 5)\ 10^{-4}$).
The order of magnitude of the critical stress
is roughly that of the low shear viscosity of the suspension
multiplied by the critical shear rate.
The Krieger-Dougherty model
\cite{Larson1999} for hard spheres: $\eta = \eta_s \left(1 -
\phi/\phi_m \right)^{-2.5 \phi_m}$, gives $\sigma_c
\mathbf{= \eta \dot{\gamma_c}} \approx 1$ Pa. This gives
the correct order of magnitude of $\sigma_c$ but also explains why
the critical stress remains constant: this follows from combining
$\dot{\gamma_c} \propto 1/\eta_s$ and $\eta \propto \eta_s$
\cite{Le_c}.

\begin{figure}
\includegraphics[width=5cm]{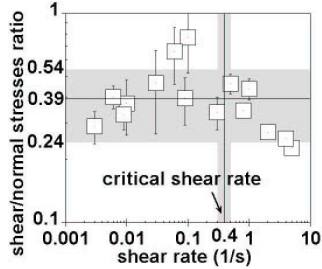}
\caption{Shear/normal stresses ratio $\sigma/N_1$ as a function of
the shear rate for polystyrene beads in 20 mPa.s silicone oil
(solid volume fraction: 58\%), at imposed shear rate. Both plates
are covered with sand paper to avoid wall slip. The shadow zone is the
statistical error bar. The horizontal line appears off center because of the logarithmic scale.
}\label{fig3}
\end{figure}

All of the rheometric experiments agree with the hypothesis
that the low-shear, low interstitial fluid viscosity regime is frictional, in the
sense that the material behaves similarly to dry granular matter.
The remaining question is whether the second regime is viscous.
Surprisingly, measurements of the first normal stress difference
$N_1$ (neglecting the second normal stress difference) using a
plate-plate geometry (diameter 40 mm) show that
the normal and the viscous stresses are proportional in both flow regimes.
For the plate-plate geometry, the critical shear rate is found
to be very similar to what found
in the Couette cell ($\dot{\gamma_c}$ = 0.5 $\pm$ 0.2 s$^{-1}$).
Fig. \ref{fig3} shows that the ratio $\sigma/N_1$ does not vary significantly
with the shear rate $\sigma/N_1 = 0.39 \pm 0.15$ in both regimes, and over $4$ decades in shear rate.
Previous results in the 'viscous' regime are consistent with our
findings \cite{Prasad1995, Bagnold1954, Hunt2002},
despite the fact that others \cite{Ancey1999}
have reported different results.
Brady and Morris \cite{Brady97} have shown theoretically
that a shear stress and a pressure proportional to shear rate
may result from hydrodynamic interactions between the particles;
however in order to generate normal stress \textit{differences},
hard sphere contacts must be incorporated. As a consequence,
there is no reason for the ratio $\sigma/N_1$ to be equal
in the frictional and lubricated regime as observed
experimentally. This surprising observation is reinforced by
the finding that the ratio $\sigma/N_1$ is equal to
the internal static friction coefficient of the dry granular
material $\mu_s = 0.40 \pm 0.04$ (as obtained from the slope angle
$\theta$ of a heap of dry beads with $\mu_s = \tan \theta$). We conclude that,
in the lubricated regime, the material
has a number of characteristics of dry granular materials, while it
simultaneously dissipates viscously in the interstitial fluid (as follows from $\eta
\propto \eta_s$).

These puzzling observations raise the following questions: what happens in the
flow? What is the distinction between the two regimes?
To address these questions, we carried out experiments with a
velocity controlled ``MRI-rheometer'', which allows for a direct
measurement of the local velocity distribution in a Couette
geometry \cite{IRM}.

\begin{figure}
\includegraphics[width=6cm]{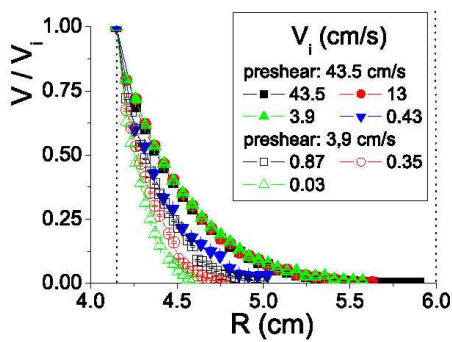}
\caption{Velocity rescaled with the velocity of the inner
cylinder
(inner cylinder radius $R_i=4.15$ cm, outer cylinder
radius $R_e =6$ cm, height 11 cm; we again use the 20 mPa.s
Rhodorsil silicone oil).
For technical reasons, $V_i$ can be varied either
between 0.01 and 3.91 cm/s, or between 0.43 and 43.5 cm/s. As in
the macroscopic rheology experiments, we preshear the material at
a rotational speed $V_i$=43.5 cm/s during 30 s. For the low
velocity setup, we preshear the material at the maximum possible
rotational speed $V_i$, i.e. 3.91 cm/s.
The rough inner cylinder is
driven at a rotational velocity $V_i$ ranging between $0.013$ and
$43.5$ cm/s, yielding overall shear rates between 0.006 s$^{-1}$
and 20.9 s$^{-1}$, the critical shear rate $\dot{\gamma_c}$ being
0.4 $\pm$ 0.1 s$^{-1}$ (or $V_c$ = 1.04 $\pm$ 0.26 cm/s).
}\label{fig4}
\end{figure}

The main results from these MRI measurements are that the velocity
profiles are roughly exponential as in dry granular materials \cite{Mueth2000},
and that they
occupy only a small fraction of the gap at low rotation rates,
i.e. in the frictional regime: we
observe shear localization. However, if the rotation
frequency is increased, a surprising behavior is observed: contrary to what
happens for dry granular matter, the higher the rotation rate, the
larger the fraction of the paste that is sheared (Fig.
\ref{fig4}). In the lubricated regime,
beyond $\dot{\gamma_c}$ = 0.4 $\pm$ 0.1 s$^{-1}$,
the whole sample is sheared ($\dot{\gamma_c}$ 
measured with the MRI is the same
as $\dot{\gamma_c}$ found in the rheology) \cite{Barentin2004}. 

\begin{figure}
\includegraphics[width=9cm]{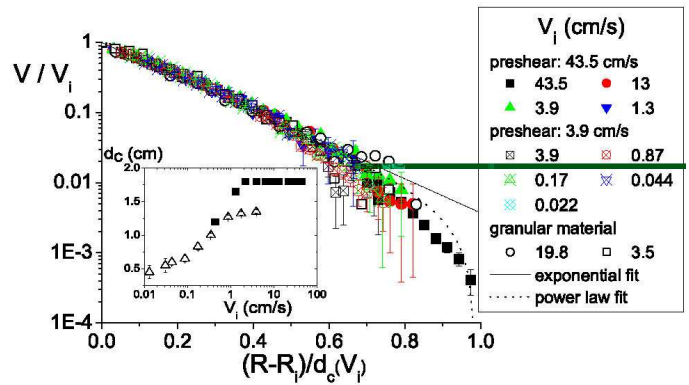}
\caption{Velocity profiles measured in a Couette geometry. We plot
$V/V_i$ vs. $(R-R_i\!)/d_c(V_i\!)$ where $R_i$ is the inner
cylinder radius and $d_c$ an adjustable parameter; we also plot
two velocity profiles extracted from \cite{Dacruz2004} for the shearing of a
dry granular material (1 mm mustard grains) in a Couette geometry
($R_i=3$ cm, $R_e=6$ cm). The line is an exponential fit: $V/V_i =
\mbox{e}^{-5.6\frac{R-R_i}{d_c(V_i\!)}}$; the dotted line is for a
power-law fluid fit with exponent $n=0.13$. Inset: $d_c$ vs. $V_i$
for a 3.91 cm/s preshear (open triangles) and a 43.5 cm/s preshear
(squares).}\label{fig5}
\end{figure}

In the frictional regime, the behavior is only
different in the sense that a smaller fraction of the material is
sheared. In this regime the reduced velocity
$V(R)/V_i$ ($V_i$ being the velocity of the rotating inner cylinder)
can be collapsed onto the same universal curve when
plotted as a function of the rescaled coordinate
$(R-R_i)/d_c(V_i)$ (Fig. \ref{fig5}). The length $d_c(V_i)$ simply
gives the extent of the material that is sheared. In addition we
find that this universal rescaling of the velocity profile also
applies to dry granular materials: two velocity profiles from \cite{Gdrmidi2004}
and \cite{Dacruz2004} fall along the same universal curve (Fig.
\ref{fig5}), underlining the universality of the roughly exponential velocity profiles
which goes beyond granular pastes. The inset to Fig. \ref{fig5} gives the extent of the
sheared region $d_c$ as a function of the rotation rate. Starting
from low $V_i$, $d_c$ increases and eventually fills the whole gap for
$V_i>V_c$ \cite{footnote_profils_vitesse_maitresse}.

This last observation provides a natural explanation for the macroscopic rheology
data. In the first, frictional regime, no steady flow without localization can be achieved
at imposed shear rates :
there is a coexistence
between a sheared and an unsheared region, which results in a
constant stress, not unlike stress plateaux and the corresponding
shear bands observed for certain surfactant and polymer solutions
\cite{Olmsted}. When the sheared region has invaded the gap, there
is no longer coexistence and the stress increases again
with increasing shear rate. The transition between the frictional
and lubricated regimes then happens at the end of the coexistence.

\begin{figure}
\includegraphics[width=6cm]{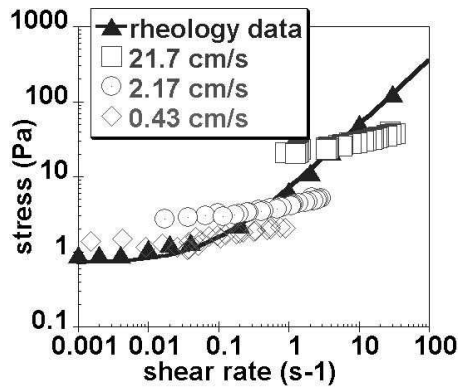}
\caption{Stress vs. shear rate calculated from the velocity
profiles for different rotational velocities and measured
macroscopically (same data as Fig. \ref{fig1}b).}\label{fig6}
\end{figure}

The interesting question is then whether a viscosity can be
defined for the material, in the lubricated regime at least, that completely
characterizes its resistance to flow. If we suppose our material
is a continuum medium in stationary flow, momentum conservation
leads to $\sigma(R)=\sigma_i R_i^2/R^2$ (where $\sigma_i$ is the
total shear stress on the inner cylinder), independently of the
constitutive equation of the material: the stress varies within
the gap. The shear rate is the spatial derivative of the
velocity profile. Therefore it varies within the gap and a single MRI
experiment allows to recalculate a complete stress-shear
rate curve. The result is shown in Fig. \ref{fig6} for different
rotation velocities.
There are two surprises: first, the data are not consistent
between different MRI experiments, and second, they are
inconsistent with the macroscopic rheology data \cite{global_local_granular}.
We are therefore forced to conclude that the velocity profiles are
not consistent with the macroscopic rheometric data, both when
there is an unsheared region in the material and when the
material is entirely sheared. The possible reasons for this is that
either the material is inhomogeneous,
or that no simple constitutive equation relating shear stress to shear rate
exists for the material (as in dry granular
materials \cite{Gdrmidi2004}).

The question is then: what happens in the lubricated regime? We can
reasonably assume that, even if force chains are
present in the sheared system, momentum conservation is likely to
hold, at least on average, therefore $\sigma \propto \sigma_i/R^{2}$. If
we combine this with the roughly exponential decay of the velocity
profile, and calculate a local viscosity from the ratio of the two,
it follows that this local viscosity is small near the moving inner
cylinder, and increases with increasing distance from the moving
wall. This is in fact easy to imagine, if the particle concentration
is slightly smaller near the wall, and increases with the
radius. The expected concentration profile can easily be evaluated
with a Krieger-Dougherty-like model for the dependence of the
viscosity on the volume fraction; it is outside of the scope of
this paper, but we plan to measure the concentration profile with the MRI and
compare them to what is expected from the velocity profiles.

In conclusion, we have studied the transition
between the frictional and lubricated flows of a dense
paste. We find that 
the bifurcation of viscosity completely and unambiguously characterizes the transition
between jammed and flowing states under imposed stress, and between localized flows and
homogeneous flows without localization under imposed shear rate.
The bifurcation gives both the critical stress and critical shear rate,
the critical shear rate being inversely
proportional to the fluid viscosity. 
We have shown that the critical Leighton number, which follows from the critical shear rate and stress and from the solvent viscosity, completely fixes the transition between the two regimes.

In the
'frictional' regime, the shear flow is localized, with a shear
zone increasing with the macroscopic shear rate unlike
what happens for dry
granular flows. Moreover, the
'lubricated' regime is not simply viscous: the
ratio of normal to shear stresses remains constant, the velocity
profiles are roughly exponential and can be rescaled in a
universal way in the two regimes. Finally we show that either the
material is inhomogeneous, or that no simple constitutive relation
exists in the lubricated regime; it is in any case impossible to
define a macroscopic viscosity that does not
depend on the measurement system or the geometry, both in the
frictional and in the lubricated regime.

We thank F. Chevoir and F. da Cruz for their data,
and A. Goriely for a critical reading of the manuscript.


\begin{thebibliography}{99}
\bibitem{silicone} The density of these oils is 0.96
g.cm$^{-3}$. Due to the small density difference we are at
a volume fraction of 58\%; this small difference turns out to be
unimportant; in addition sedimentation effects are small due to
the fact that we are at high solid loadings.
\bibitem{infinite_viscosity} The apparent saturation at a finite value of the viscosity
is due to the finite resolution of the position transducer of the rheometer.
\bibitem{footnote_profils_vitesse_maitresse} $d_c$ is slightly smaller for a 9 rpm preshear.
This is likely to be a sedimentation effect, since resuspension is
more efficient at 100 rpm. However, this does not affect the
generality of our results; rheometric experiments performed in
both conditions show very little difference.
\end{thebibliography}
\end{document}